\documentclass{article}
\usepackage{amssymb}
\usepackage{amsmath}
\usepackage{array}
\usepackage{algorithm}
\usepackage{algpseudocode}
\usepackage{booktabs}
\usepackage{color}
\usepackage{dcolumn}
\newcolumntype{d}[1]{D{.}{.}{#1}}
\usepackage[hang,flushmargin]{footmisc}
\usepackage{lipsum}
\usepackage{multirow}
\usepackage[numbers,sort&compress]{natbib}
\usepackage{spconf,graphicx}
\usepackage{subfigure}
\usepackage{bm}
\usepackage{siunitx}
\usepackage{subcaption} 
\usepackage{threeparttable}

\title{Robust Online Overdetermined Independent Vector Analysis Based on Bilinear Decomposition}

\name{
	\begin{tabular}{c}
		Kang Chen$^1$, Xianrui Wang$^{2,6}$, Yichen Yang$^{2,6}$, Andreas Brendel$^3$, \\
		Gongping Huang$^1$, Zbyněk Koldovský$^4$, Jingdong Chen$^2$, Jacob Benesty$^5$, and Shoji Makino$^6$
	\end{tabular}
	\thanks{This work was supported by the National Key Research and Development Program of China under Grant No. 2021ZD0201502, the National Natural Science Foundation (NSFC) of China under Grant 62471340, 62192713, and 61831019, and the Czech Science Foundation (GA\v{C}R) No. 25-18485S.}
}
\address{ \large \hskip -7pt \begin{tabular}{c}
		$^1$School of Electronic Information, Wuhan University, 430072, Wuhan, China\\
		$^2$CIAIC and Shaanxi Provincial Key Laboratory of Artificial Intelligence, \\
Northwestern Polytechnical University, Xi’an, Shaanxi, China \phantom{Intelli}\\
		$^3$Fraunhofer IIS, Fraunhofer Institute for Integrated Circuits IIS, Erlangen, Germany  \\
		$^4$Technical University of Liberec, Liberec, Czech Republic \\
		$^5$INRS-EMT, University of Quebec, Montreal, Canada \\
		$^6$Waseda University, Japan
	\end{tabular}
}

\begin{document}
	
	\ninept
	\maketitle
	
\begin{abstract}
Online blind source separation is essential for both speech communication and human–machine interaction. Among existing approaches, overdetermined independent vector analysis (OverIVA) delivers strong performance by exploiting the statistical independence of source signals and the orthogonality between source and noise subspaces. However, when applied to large microphone arrays, the number of parameters grows rapidly, which can degrade online estimation accuracy. To overcome this challenge, we propose decomposing each long separation filter into a bilinear form of two shorter filters, thereby reducing the number of parameters. Because the two filters are closely coupled, we design an alternating iterative projection algorithm to update them in turn. Simulation results show that, with far fewer parameters, the proposed method achieves improved performance and robustness.
\end{abstract}


\section{Introduction}
\label{Sect-Intro}

In real-world environments, multiple sound sources are often mixed together, which not only reduces the quality of speech communication but also impairs the performance of human–machine interaction~\cite{benesty2023microphone, huang2025advances, gannot2017consolidated}. The problem becomes even more challenging when neither source position priors nor source activity information is available. To address this, blind source separation~(BSS), which aims to recover source signals without relying on such prior knowledge, has drawn significant research attention, leading to the development of numerous algorithms~\cite{Belouchrani1998Blind,Bofill2001Underdetermined,Makino2018audio,Makino2007Blind,Buchner2004TRINICON}. Among these, independent vector analysis (IVA) has become one of the most widely adopted, which separates source signals by leveraging their statistical independence~\cite{Kim2006independent, Brendel2021Fasteriva, Wang2023Spatially, jansky2022auxiliary, ono2012auxiliary}. The majorization–minimization (MM) framework was later introduced into IVA, giving rise to the well-known auxiliary-function-based variant, AuxIVA~\cite{Ono2011Stable}. However, classical IVA assumes a determined mixing model, where the number of microphones equals the number of sources. This assumption prevents full utilization of spatial diversity when an array contains more microphones than sources. To address this issue, researchers proposed the orthogonal constraint (OC), which enforces orthogonality among the subspaces of separated signals~\cite{koldovsky2017orthogonally, koldovsky2018gradient, Ikeshita2020Overdetermined}. Incorporating OC yields overdetermined IVA (OverIVA), which has been demonstrated to outperform AuxIVA in various scenarios~\cite{Scheibler2019Independent, Du2022Computationally, Ikeshita2020Overdetermined}.

However, the number of parameters in OverIVA grows rapidly with the array size, which leads to biased statistical estimates, particularly in online implementations. Consequently, OverIVA becomes unsuitable for real-time communication and low-latency human–machine interaction when large microphone arrays are employed. To overcome this limitation, we propose enhancing the online separation performance of OverIVA by applying a bilinear decomposition of the demixing filters into much shorter sub-filters~\cite{huang2020kronecker,Paleologu2018Linear,Benesty2019Array,Wang2022A,Wang2021Time}. To effectively estimate the sub-filters under strong coupling, we introduce an alternating iterative projection algorithm for their updates~\cite{wang2024semi}. This bilinear decomposition drastically reduces the number of filter coefficients, thereby enhancing robustness and improving separation performance over conventional OverIVA, as confirmed by extensive simulations.

The remainder of this paper is organized as follows. Section~\ref{signal_model} introduces the signal model, problem formulation, and provides a brief review of the OverIVA algorithm. The proposed online OverIVA method is derived in Section~\ref{sec-proposed}. Section~\ref{sec-simulation} presents simulation results and analysis, and Section~\ref{sec-conclusion} concludes the paper.

\section{Signal model}
\label{signal_model}

\subsection{Problem Formulation}
We consider a noisy and reverberant overdetermined acoustic scenario in which $N$ target source signals are captured by $M$ microphones ($M > N$). In the short-time Fourier transform (STFT) domain, the microphone signal vector at frequency bin $i$ and time frame $j$, defined as 
$\mathbf{x}_{i,j} = \left[X_{1, i, j}~~X_{2, i, j}~~\ldots~~X_{M,i,j}\right]^T \in \mathbb{C}^M$, can be represented as
\begin{equation}
	\mathbf{x}_{i,j}=\mathbf{A}_{i,j}\mathbf{s}_{i,j}+\mathbf{\Psi}_{i,j}\mathbf{z}_{i,j},
	\label{Freq_Mat_Mix}
\end{equation}
where $\mathbf{s}_{i,j} = \left[S_{1, i, j}~~S_{2, i, j}~~\ldots~~S_{N,i,j}\right]^T \in \mathbb{C}^N$ is the source signal vector, $\mathbf{z}_{i,j} = \left[Z_{1, i, j}~~Z_{2, i, j}~~\ldots~~Z_{M-N,i,j}\right]^T \in \mathbb{C}^{M-N}$ contains the noise, and  $\mathbf{A}_{i,j} \in \mathbb{C}^{M \times N}$ and $\mathbf{\Psi}_{i,j} \in \mathbb{C}^{M \times M-N}$ are source mixing and noise mixing matrices, respectively. Both $\mathbf{A}_{i,j}$ and $\mathbf{\Psi}_{i,j}$ are time-dependent, reflecting the online scenario where the mixing may vary over time. The goal of online BSS is to recover the source signals from the microphone observations, $\mathbf{x}_{i,j}$, by applying a demixing matrix:
\begin{equation}
	\mathbf{W}_{i,j}=\left[\begin{array}{c}
		\mathbf{W}_{i,j}^{\mathrm{s}} \\[5pt]
		\mathbf{U}_{i,j}
	\end{array}\right] \in \mathbb{C}^{M \times M},
	\label{Demix_Mat_Global}
\end{equation}
where 
\begin{align}
	\mathbf{W}_{i,j}^{\mathrm{s}}=\left[\mathbf{w}_{1,i,j} ~~ \mathbf{w}_{2,i,j} ~~ \ldots ~~ \mathbf{w}_{N,i,j}\right]^H \in \mathbb{C}^{N \times M}
\end{align}
and $\mathbf{U}_{i,j} \in \mathbb{C}^{M-N \times M}$ are the source extraction matrix and noise separation matrix, respectively. By applying the source separation filters to the observed signals, the target signals can be extracted as
\begin{eqnarray}
	Y_{n,i,j} = \mathbf{w}^H_{n,i,j}\mathbf{x}_{i,j},
	\label{Source_Demix_Process}
\end{eqnarray}
where $Y_{n,i,j}$ is the estimate of the $n$th source signal. 

\subsection{Online OverIVA}
Most existing BSS approaches assume that source signals follow either a non-Gaussian distribution, such as a circularly symmetric Laplace distribution or a time-varying complex Gaussian distribution~\cite{Ono2011Stable, koldovsky2022double, Scheibler2019Independent}. In this work, we adopt the latter, whose probability density function is given by $p\left(\mathbf{y}_{n,j}\right) \propto  \exp\left(-{\vert \vert \mathbf{y}_{n,j} \vert\vert^2_2}{/}{\sigma^2_{n,j}}\right)$, where $\mathbf{y}_{n,j} = \left[Y_{n,1,j}~~ Y_{n,2,j}~~\cdots~~Y_{n,I,j}\right]^T \in \mathbb{C}^{I}$, $\sigma^2_{n,j}$ is the broadband time-varying variance, and {$\vert\vert\cdot\vert\vert_2$} denotes the $\ell_2$ norm. The noise components, in contrast, are modeled as time-invariant complex Gaussian variables and are estimated using the orthogonal constraint between the subspaces of separated sources and noise. In OverIVA, this orthogonality prevents noise from directly interfering with the estimation of the target signals. Consequently, the objective function for estimating the time-varying demixing matrices $\mathcal{W} = \{\mathbf{W}_{i,j}^{\mathrm{s}}\}_{i=1}^I$ at time frame $j$ can be formulated as~\cite{taniguchi2014auxiliary, nakashima2023fast}
\begin{align}
	\nonumber
	\mathcal{J}(\mathcal{W}) = \frac{1}{\sum_{t=1}^j \alpha^{j-t}} \sum_{t=1}^{j}\sum_{n=1}^{N}\alpha^{j-t}&\frac{{\vert \vert \mathbf{y}_{n,t} \vert\vert^2_2}}{\sigma^2_{n,t}} \\
	&- 2\sum_{i=1}^I \log |\det \mathbf{W}_{i,j}|,
	\label{Negated_Likelihood_All_Global}
\end{align}
where $\alpha \in (0,1]$ is a forgetting factor and $t$ denotes the {past} frame index. 
Since the cost function is {difficult} to optimize directly, {the} auxiliary function technique is utilized {by optimizing the upper bound}, i.e., 
\begin{align}\label{Aux_W_Global}
	\mathcal{J}^+(\mathcal{W}) = \sum_{i=1}^{I}\sum_{n=1}^{N}\mathbf{w} _{n,i,j}^H\mathbf{V}_{n,i,j}\mathbf{w}_{n,i,j}- 2\sum_{i=1}^I \log |\det \mathbf{W}_{i,j}|,
\end{align}
where
\begin{align}
	\mathbf{V}_{n,i,j} &= \alpha \mathbf{V}_{n,i,j-1} + (1-\alpha)\phi_{\mathrm{y}_{n,j}}\mathbf{x}_{i,j}\mathbf{x}_{i,j}^H,\\
	\label{phi_n_j}
	\phi_{\mathrm{y}_{n,j}} &= \frac{1}{\sum_{i=1}^{I}\left| \mathbf{w}_{n,i,j}^H\mathbf{x}_{i,j}\right|^2}
\end{align}
are the weighted covariance matrix  and the weighting function derived from the target source distribution, respectively. 
Since $\phi_{\mathrm{y}_{n,j}}$  is frequency independent, \eqref{Aux_W_Global} can be decomposed into separate components for each frequency. Hence, the frequency index $i$ is omitted in the following discussion unless explicitly required. By differentiating \eqref{Aux_W_Global} and applying the iterative projection method, the rows of  $\mathbf{W}_{j}^{\mathrm{s}}$ are updated as~\cite{Ono2011Stable, ono2012auxiliary}
\begin{align}
	& \mathbf{w} _{n,j}\ {\leftarrow}\ \left(\mathbf{W} _{j} \mathbf{V}_{n,j}\right)^{-1} \mathbf{e}_n, \\
	& \mathbf{w} _{n,j}\ {\leftarrow}\ \frac{\mathbf{w} _{n,j}}{\sqrt{\mathbf{w} _{n,j}^{\mathrm{H}} \mathbf{V}_{n,j} \mathbf{w} _{n,j}}},
\end{align}
where $\mathbf{e}_n$ denotes the unit vector whose $n$th entry is equal to one.

After estimating all source extraction filters, the noise separation matrix $\mathbf{U}_{j}$ must be updated. Since the goal is not to separate individual noise components but only to distinguish them from the target signals, $\mathbf{U}_{j}$ can be structured as $\mathbf{U}_{j}=\left[\mathbf{J}_{j}~~ -\mathbf{I}_{M-N} \right]$, where $\mathbf{J}_{j}{\in\mathbb{C}^{M-N \times N}}$ {contains} adjustable parameters and $\mathbf{I}_{M-N}$ is the $(M-N) \times (M-N)$ identity matrix. By applying the orthogonal constraint between noise and target subspaces, $\mathbf{J}_{j}$ can be derived as
\begin{equation}
	\begin{aligned}
		\mathbf{J}_j := \left\{ \mathbf{R}_N \mathbf{C}_j \left(\mathbf{W}^{\mathrm{s}}_j\right)^H \right\} 
		\left\{\mathbf{R}_S \mathbf{C}_j\left(\mathbf{W}^{\mathrm{s}}_j\right)^H\right\}^{-1} ,
	\end{aligned}
	\label{J_Update}
\end{equation}
where 
\begin{align}
		\mathbf{C}_{j} &= \alpha \mathbf{C}_{j-1} + (1-\alpha)\mathbf{x}_{j}\mathbf{x}_{j}^H \label{C_frame}
\end{align}
is the recursively estimated spatial covariance matrix of the observed signals, with $\mathbf{R}_N = \left[ \mathbf{0}_{M-N\times N} \quad  \mathbf{I}_{M-N}  \right]$ and $\mathbf{R}_S = \left[\mathbf{I}_N\quad  \mathbf{0}_{N\times M-N}\right]$ \cite{Scheibler2019Independent}.

\section{Proposed Method: Online OverIVA with Bilinear Decomposition}
\label{sec-proposed}
In this section, we propose decomposing each source extraction filter into a bilinear form, where the original long filter is expressed as the Kronecker product of two shorter sub-filters~\cite{Paleologu2018Linear,Benesty2019Array,Wang2022A}. Let $M = M_1 M_2$, with the assumption, without loss of generality, that $M_1 \geq M_2$. Under this formulation, the source extraction filter can be decomposed as
\begin{equation}
	\mathbf{w}_{n,j} = \mathbf{w}_{n,j,1}\otimes \mathbf{w}_{n,j,2},
	\label{Kron_Decomp}
\end{equation}
where $\otimes$ denotes the Kronecker product, and  $\mathbf{w}_{n,j,1}$ and $\mathbf{w}_{n,j,2}$ are two sub-filters of length $M_1$ and $M_2$, respectively. For simplicity, we set $\vert\vert \mathbf{w}_{n,j,1} \vert\vert_2 = 1$, since scaling $\mathbf{w}_{n,j,1}$ and inversely scaling $\mathbf{w}_{n,j,2}$ does not affect \eqref{Kron_Decomp}. Because $M_1 M_2 > M_1 + M_2$, the bilinear decomposition requires fewer parameters than the original filters. This reduction enhances robustness, particularly as the number of microphones increases.

We then have the following relationship:
\begin{equation}
	\begin{aligned}
	\label{Kron1_Rela}
	\mathbf{w}_{n,j} &= \mathbf{w}_{n,j,1}\otimes \mathbf{w}_{n,j,2}\\
	&= \left(\mathbf{I}_{M_1} \otimes \mathbf{w}_{n,j,2} \right) \mathbf{w}_{n,j,1} \\
	&= \Delta_{n,j,1}\mathbf{w}_{n,j,1},
	\end{aligned}
\end{equation}
where $\mathbf{I}_{M_1}{\in\mathbb{R}^{M_1\times M_1}}$ is the identity matrix and $\Delta_{n,j,1} = \mathbf{I}_{M_1} \otimes \mathbf{w}_{n,j,2}{\in\mathbb{C}^{M \times M_1}}$. {Similarly}, \eqref{Kron_Decomp} can {be rewritten} as
\begin{equation}
	\begin{aligned}
		\label{Kron2_Rela}
		\mathbf{w}_{n,j} &= \mathbf{w}_{n,j,1}\otimes \mathbf{w}_{n,j,2}\\
		&= \left( \mathbf{w}_{n,j,1} \otimes  \mathbf{I}_{M_2} \right) \mathbf{w}_{n,j,2} \\
		&= \Delta_{n,j,2}\mathbf{w}_{n,j,2},
	\end{aligned}
\end{equation}
where $\mathbf{I}_{M_2}{\in\mathbb{R}^{M_2\times M_2}}$ is the identity matrix and $\Delta_{n,j,2} = \mathbf{w}_{n,j,1} \otimes  \mathbf{I}_{M_2}{\in\mathbb{C}^{M \times M_2}}$. 

With the above decomposition, the estimation of the demixing matrices $\mathcal{W}$ is reduced to estimating two sets of shorter sub-filters, $\mathcal{W}_{1}$ and $\mathcal{W}_{2}$, corresponding to the first and second groups of sub-filters, respectively. Since the sub-filters are interdependent, we adopt an alternating iterative projection method to optimize them, extending the approach in~\cite{wang2024semi}. In fact, the method in~\cite{wang2024semi} can be viewed as a special case of our algorithm for the single-source scenario. Specifically, we first fix ${\mathbf{w}}_{n,j,2}$ and compute ${\Delta}_{n,j,1}$. Substituting \eqref{Kron1_Rela} into \eqref{Aux_W_Global} and omitting the frequency index, the cost function for ${\mathbf{w}}_{n,j,1}$ is then obtained as
\begin{align}
	\mathcal{J}^+(\mathcal{W}_1\vert  \mathcal{W}_2)
	 =\sum_{n=1}^{N}{\mathbf{w}}_{n,j,1}^H\underline{\mathbf{V}}_{n,j,1}{\mathbf{w}}_{n,j,1}  -2\log\vert\det\widetilde{\mathbf{W}}_j\vert,
	\label{Cost_SOI_w1}
\end{align}
where 
\begin{align}
	\label{V_nj1}
	\underline{\mathbf{V}}_{n,j,1} 
	={\Delta}_{n,j,1}^H \mathbf{V}_{n,j}{\Delta}_{n,j,1}
\end{align}
and $\widetilde{\mathbf{W}}_j$ denotes the source separation {matrix defined analogously to ${\mathbf{W}}_j$ but with source extraction filters in} bilinear form.
By setting the Wirtinger derivative of \eqref{Cost_SOI_w1} with respect to ${\mathbf{w}}_{n,i,1}^*$ (where $^*$ {denotes the complex conjugate}) {to $0$}, we obtain
\begin{equation}
	\underline{\mathbf{V}}_{n,j,1}{\mathbf{w}}_{n,j,1}
	={\Delta}_{n,j,1}^H\widetilde{\mathbf{W}}_j^{-1}\mathbf{e}_n.
	\label{Gradient_sub1}
\end{equation}

However, since $\widetilde{\mathbf{W}}_j$ is not directly accessible, we follow the strategy used in AuxIVA, ILRMA, and OverIVA, replacing the $n$th source extraction filter with the updated $\mathbf{w}_{n,j}$ after each iteration. The corresponding update rule for ${\mathbf{w}}_{n,j,1}$ can then be expressed as
\begin{equation}
	{\mathbf{w}}_{n,j,1} = \underline{\mathbf{V}}_{n,j,1}^{-1}{\Delta}_{n,j,1}^H {\widetilde{\mathbf{W}}_j}^{-1}\mathbf{e}_n.
	\label{Unorm_update1}
\end{equation}
Pre-multiplying ${\mathbf{w}}_{n,j,1}^H$ to both sides of \eqref{Gradient_sub1}, we obtain
\begin{align}
	{\mathbf{w}}_{n,j,1}^H\underline{\mathbf{V}}_{n,j,1}{\mathbf{w}}_{n,j,1}=1.
\end{align}
Therefore, after updating ${\mathbf{w}}_{n,j,1}$, the following normalization should be carried out
\begin{align}
	{\mathbf{w}}_{n,j,1} {\leftarrow} \frac{\mathbf{w}_{n,j,1}}{\sqrt{{\mathbf{w}}_{n,j,1}^H
		\underline{\mathbf{V}}_{n,j,1}{\mathbf{w}}_{n,j,1}}}.
	\label{Norm_w1}
\end{align}
The update rules for ${\mathbf{w}}_{n,j,2}$ are derived similarly, using the updated ${\mathbf{w}}_{n,j,1}$ {in} \eqref{Kron2_Rela} and {compute} ${\Delta}_{n,j,2}$. The corresponding cost function for ${\mathbf{w}}_{n,j,2}$ is then given by
\begin{align}
\mathcal{J}^+(\mathcal{W}_2\vert  \mathcal{W}_1)
 =\sum_{n=1}^{N}{\mathbf{w}}_{n,j,2}^H\underline{\mathbf{V}}_{n,j,2}{\mathbf{w}}_{n,j,2} -2\log\vert\det\widetilde{\mathbf{W}}_j\vert,
	\label{Cost_SOI_w2}
\end{align}
where 
\begin{align}
	\underline{\mathbf{V}}_{n,j,2} 
	&={\Delta}_{n,j,2}^H \mathbf{V}_{n,i}{\Delta}_{n,j,2}.
\end{align}
Similarly, the update rule for ${\mathbf{w}}_{n,j,2}$ can be derived as
\begin{equation}
	{\mathbf{w}}_{n,j,2} = \underline{\mathbf{V}}_{n,j,2}^{-1}{\Delta}_{n,j,2}^H{\widetilde{\mathbf{W}}}_j^{-1}\mathbf{e}_n.
	\label{Unorm_update2}
\end{equation}
The corresponding normalization is then applied as
\begin{align}
	{\mathbf{w}}_{n,j,2} {\leftarrow} \frac{{\mathbf{w}}_{n,j,2}}{\sqrt{{\mathbf{w}}_{n,j,2}^H
		\underline{\mathbf{V}}_{n,j,2}{\mathbf{w}}_{n,j,2}}}.
	\label{Norm_w2}
\end{align}

After updating all sub-filters, they are combined to form the source separation matrix. The noise separation matrix is then updated using the same procedure described in~\eqref{J_Update}. We refer to this proposed online overdetermined IVA method with bilinear decomposition as BiIVA. The complete algorithmic procedure is summarized in Algorithm~\ref{Algo_BiIVA}.

\begin{algorithm}[t!]
	\caption{Online BiIVA}
	\begin{algorithmic}[1]
		\Statex \textbf{Input:}  $N$, $\alpha$, $M_1$, $M_2$, $\mathbf{x}_{j}  $, $\forall i,j$
		\Statex  \textbf{Initialize:} ${\mathbf{w}}_{n,0,1} $, ${\mathbf{w}}_{n,0,2}$, $\widetilde{\mathbf{W}}_{i,0}$, $\mathbf{C}_{i,0}$, $\mathbf{V}_{n,i,0}$, $\forall n,i$
		\Statex \textbf{Output:} $\mathbf{y}_{j} ~ \forall i,j$
		\For{$j = 1, 2, \cdots, J$}
		\State $\widetilde{\mathbf{W}}_{j} = \widetilde{\mathbf{W}}_{j-1}$, $\forall i$
		\For{$n = 1, 2, \cdots, N$}
		\State $\phi_{\mathrm{y}_{n,j}} = {1} /{\sum_{i=1}^{I}\left(\mathbf{w}_{n,i,j}^H\mathbf{x}_{i,j}\right)^2}$
		\State 	$\mathbf{V}_{n,j} = \alpha \mathbf{V}_{n,j-1} + (1-\alpha)\phi_{\mathrm{y}_j}\mathbf{x}_{j}\mathbf{x}_{j}^H$, $\forall i$
		\State $	\mathbf{C}_{j} = \alpha \mathbf{C}_{j-1} + (1-\alpha)\mathbf{x}_{j}\mathbf{x}_{j}^H$, $\forall i$
		
		\State ${\Delta}_{n,j,1} = \mathbf{I}_{M_1} \otimes \mathbf{w}_{n,j,2} $ , $\forall i$
		\vspace{1pt}
		\State $\underline{\mathbf{V}}_{n,j,1}={\Delta}_{n,j,1}^H \mathbf{V}_{n,j}{\Delta}_{n,j,1}$ , $\forall i$
		\State 	${\mathbf{w}}_{n,j,1} = \underline{\mathbf{V}}_{n,j,1}^{-1}{\Delta}_{n,j,1}^H {\widetilde{\mathbf{W}}_j}^{-1}\mathbf{e}_n$, $\forall i$
		\State ${\mathbf{w}}_{n,j,1} = {\mathbf{w}_{n,j,1}}/{\sqrt{{\mathbf{w}}_{n,j,1}^H
				\underline{\mathbf{V}}_{n,j,1}{\mathbf{w}}_{n,j,1}}}$, $\forall i$
		\State ${\Delta}_{n,j,2} =  \mathbf{w}_{n,j,1} \otimes  \mathbf{I}_{M_2} $ , $\forall i$
		\State $\underline{\mathbf{V}}_{n,j,2}={\Delta}_{n,j,2}^H \mathbf{V}_{n,j}{\Delta}_{n,j,2}$ , $\forall i$
		\State 	${\mathbf{w}}_{n,j,2} = \underline{\mathbf{V}}_{n,j,2}^{-1}{\Delta}_{n,j,2}^H {\widetilde{\mathbf{W}}_j}^{-1}\mathbf{e}_n$, $\forall i$
		\State ${\mathbf{w}}_{n,j,2} = {\mathbf{w}_{n,j,2}}/{\sqrt{{\mathbf{w}}_{n,j,2}^H
				\underline{\mathbf{V}}_{n,j,2}{\mathbf{w}}_{n,j,2}}}$, $\forall i$
		\EndFor
		\State $\mathbf{J}_j = \left\{ \mathbf{R}_N \mathbf{C}_j \left(\mathbf{W}^{\mathrm{s}}_j\right)^H \right\} 
		\left\{\mathbf{R}_S \mathbf{C}_j\left(\mathbf{W}^{\mathrm{s}}_j\right)^H\right\}^{-1}$
		\State Compute $\widetilde{\mathbf{W}}_{j}, \forall i$ as in \eqref{Demix_Mat_Global}
		\State $\mathbf{y}_{j} = \widetilde{\mathbf{W}}_{j} \mathbf{x}_{j}$, $\forall i$
		\EndFor
	\end{algorithmic}
	\label{Algo_BiIVA}
\end{algorithm}

\section{{Simulations}}
\label{sec-simulation}
\subsection{Simulation Setup}
\label{subSect-Exp-setup}
We consider a uniform planar rectangular microphone array consisting of $36$ microphones, arranged in a $6 \times 6$ grid with an inter-element spacing of $6$~cm along both axes. Clean speech signals from {the} CMU\_Arctic dataset~\cite{kominek2004cmu} with a sampling rate of $16$~kHz are used to synthesize mixture signals. We employ the image-source model to generate room impulse responses (RIRs)~\cite{Allen1979Image,pan2023image}, with wall reflection coefficients set to achieve a reverberation time $T_{60}$ of approximately 200~ms. The microphone array is centered on the horizontal plane at a fixed height of 3~m. Two target sources are placed at coordinates (7.0, 6.0, 1.75)~m and (6.5, 6.9, 1.75)~m within a Cartesian system with the origin at the room’s bottom-left corner. Five male and five female speakers are randomly selected from the clean dataset, with each speaker’s signals concatenated to $30$-second segments. These signals are first convolved with the corresponding RIRs and then mixed to create clean mixtures with an input signal-to-interference ratio (iSIR) of $0$~dB.

\begin{figure}[t!]
	\vspace{6pt}
	\centering
	\includegraphics[width=\columnwidth, trim=5 0 0 10, clip]{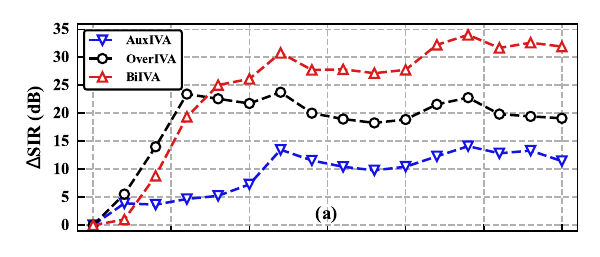}
	\includegraphics[width=\columnwidth, trim=5 0 0 10, clip]{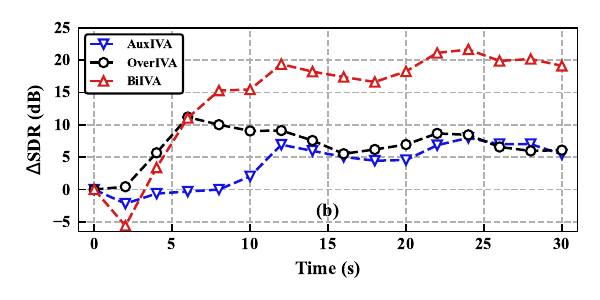}
	\caption{Convergence performance of the compared source separation algorithms: (a) SIR improvement and (b) SDR improvement. Scores are computed over $2$-second segments and averaged across $50$ samples and two speakers.}
	\label{Convergence_Behave_sir_sdr}
\end{figure}

\begin{figure}[t!]
	\centering
	\includegraphics[width=\columnwidth, trim=10 10 10 10, clip]{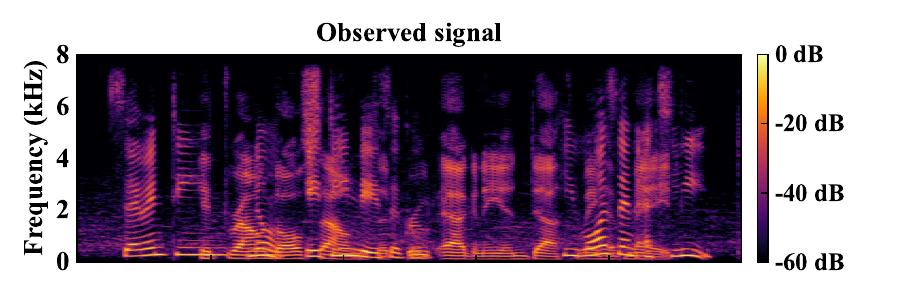}
	\includegraphics[width=\columnwidth, trim=10 10 10 10, clip]{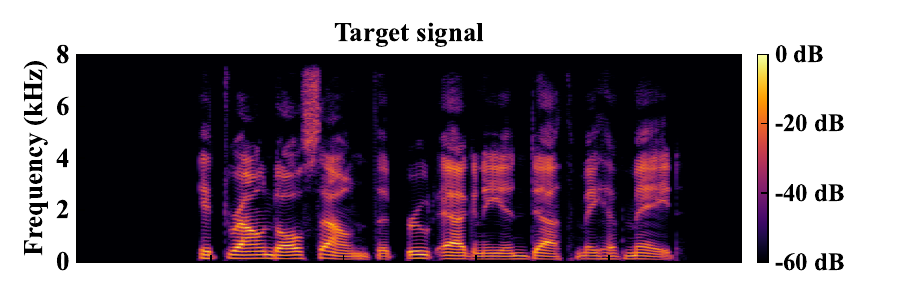}
	\includegraphics[width=\columnwidth, trim=10 10 10 10, clip]{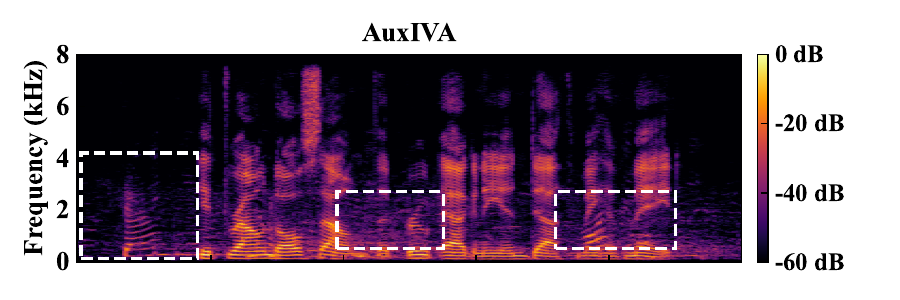}
	\includegraphics[width=\columnwidth, trim=10 10 10 10, clip]{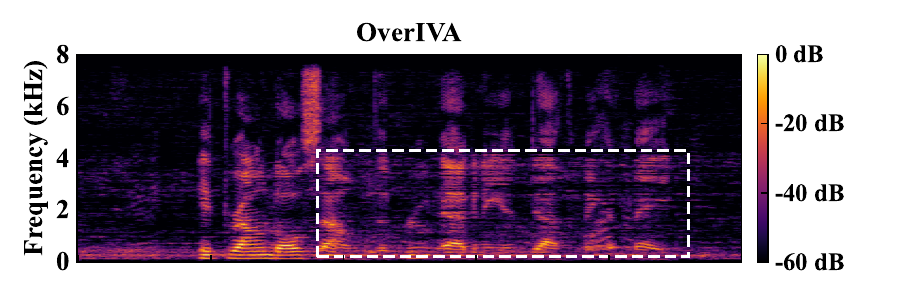}
	\includegraphics[width=\columnwidth, trim=10 10 10 10, clip]{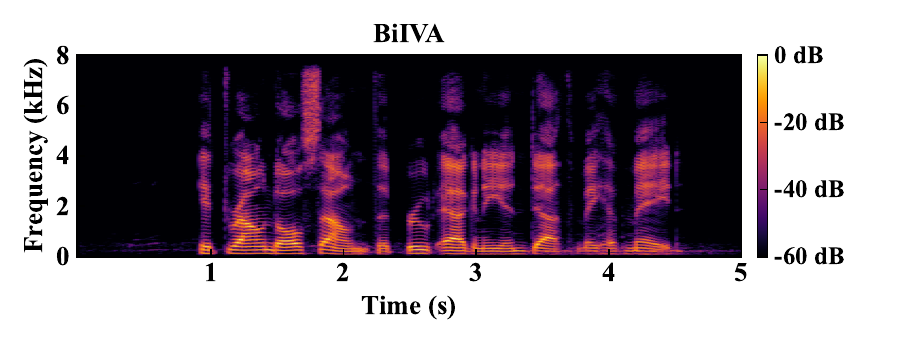}
	\caption{Spectrograms of the observed signal, the target signal, and the separated signals obtained using AuxIVA, OverIVA, and the proposed BiIVA method.}
	\label{Spectrograms}
\end{figure}

To generate the noisy observations, noise is added to the clean, convolutive mixtures. Five point noise sources are randomly placed in the room with the constraints that each is at least $0.5$~m from the walls, at least $3$~m from the room center in the $x$-$y$ plane, and separated from adjacent sources by at least $20^\circ$ in the $x$-$y$ plane. Random clips of matching length are extracted from an hour-long recorded office noise signal and convolved with the corresponding RIRs at the noise source positions. In addition, multichannel white Gaussian noise of equal variance is added to model microphone imperfections. Let $v_{\mathrm{p},m}(t)$ and $v_{\mathrm{w},m}(t)$ denote the observations of the point noise sources and white noise, respectively, at the $m$th microphone. The overall multichannel noise is then given by $v_m(t) = \sigma_{v}[v_{\mathrm{p},m}(t)+10^{-0.75}v_{\mathrm{w},m}(t)]$, where $\sigma_{v}$ is an adjustable parameter controlling the input signal-to-noise ratio (iSNR). The amplitude of each source is scaled to achieve a target iSNR of $20$~dB.

The algorithms are implemented in the STFT domain with a 1024-point Hann window and $75\%$ overlap between neighboring frames. For the proposed BiIVA method, we set $M_1 = 6$ and $M_2 = 6$, initialize $\mathbf{w}_{n,0,1} = \mathbf{e}_n$ and $\mathbf{w}_{n,0,2} = \mathbf{e}_1$ for $\forall n,i$. For online AuxIVA, the first two channels are used for separation, while the demixing matrices for both AuxIVA and OverIVA are initialized as identity matrices. The forgetting factors $\alpha$ for online BiIVA, AuxIVA, and OverIVA are set to 0.98, 0.96, and 0.99, respectively, corresponding to the best observed performance for each algorithm. Separation performance is evaluated using SIR and SDR, with iSIR, iSNR, SIR, and SDR measured at the first microphone. All simulations use 50 randomly selected signal instances to compute average performance. When extracting one target signal, all other sources are treated as interfering sources.

\subsection{Results and Analysis}
\label{subSect-result}
Figure~\ref{Convergence_Behave_sir_sdr} presents the average SIR and SDR improvements of all compared algorithms over $50$ samples and two speakers. AuxIVA shows limited improvement, with SDR reaching approximately $8$~dB and SIR up to $14$~dB after convergence, and exhibits the slowest convergence. In contrast, OverIVA delivers better separation, with SIR rising rapidly to around $20$~dB within 5 seconds and SDR steadily improving to about $10$~dB, indicating enhanced separation and reconstruction compared to AuxIVA. Notably, BiIVA outperforms all the other compared methods, achieving an SDR peak near $20$~dB and SIR exceeding $30$~dB. While its early-stage performance is slightly below OverIVA, BiIVA ultimately attains the highest overall signal quality and interference suppression after convergence. Figure~\ref{Spectrograms} further illustrates these differences through spectrograms of processed signal segments. AuxIVA and OverIVA introduce varying levels of distortion, with AuxIVA showing weaker interference suppression and OverIVA noticeable distortion (highlighted within the white box). In contrast, BiIVA effectively preserves the target signal while strongly suppressing interference, demonstrating its effectiveness.

\section{Conclusions}
\label{sec-conclusion}

This paper introduced an improved online variant of OverIVA designed to enhance performance in large microphone array scenarios through bilinear decomposition. In the proposed approach, each long separation filter is factorized into the Kronecker product of two much shorter sub-filters, substantially reducing the number of parameters to estimate. To address the interdependence between these sub-filters, an alternating iterative projection algorithm was developed for their updates. Simulations based on the CMU\_Arctic dataset, real recorded noise, and image-model-generated room impulse responses demonstrate that the proposed BiIVA method consistently outperforms conventional AuxIVA and OverIVA, yielding approximately $10$-dB improvements in both SIR and SDR after convergence. These results underscore the effectiveness of the method and its potential for practical use in overdetermined scenarios.

\end{document}